\documentclass{ametsoc}

\journal{jamc}

\bibpunct{(}{)}{;}{a}{}{,}

\title{Performance of a simple proxy for U.S. cloud-to-ground lightning}

\authors{Michael K.\ Tippett\correspondingauthor{Dept.\ of Applied
    Physics and Applied Mathematics, Columbia University, 500 West
    120th St., New York, NY/USA.}}

\affiliation{Department of Applied Physics and Applied Mathematics,
  Columbia University, New York, New York}

\email{mkt14@columbia.edu}

\extraauthor{Chiara Lepore}
\extraaffil{\vspace{-0.80em}Lamont-Doherty Earth Observatory, Columbia University, Palisades, New York}

\extraauthor{William J. Koshak}
\extraaffil{\vspace{-0.80em}NASA Marshall Space Flight Center, Huntsville, Alabama}

\extraauthor{Themis Chronis}
\extraaffil{\vspace{-0.80em}University of Alabama in Huntsville, Huntsville, Alabama}

\extraauthor{Brian Vant-Hull}
\extraaffil{\vspace{-0.80em}City College of New York, New York City, New York}

\abstract{The product of convective available potential energy (CAPE)
  and precipitation rate has previously been used as a proxy for
  cloud-to-ground (CG) lightning flash counts in climate change
  applications. Here the ability of this proxy, denoted CP, to
  represent the climatology and variability of CG lightning flash
  counts over the contiguous U.S. (CONUS) during the period 2003--2016
  is assessed.  CP values computed using the North American Regional
  Reanalysis are compared with negative and positive polarity CG flash
  counts from the National Lightning Detection Network.  Overall, the
  proxy performs better on shorter time scales (daily and monthly)
  than on longer time scales (annual and semi-annual). Proxy
  performance tends to be worse during the warm season (May--October),
  when most lightning occurs, and better during the cool season
  (November--April).
  The correlation of annually accumulated CONUS CP with CG flash
  counts is not statistically significant because of poor warm-season
  performance.  Cool season negative CG flash counts are
  well-correlated with CONUS CP values.  Positive CG flash counts
  ($\sim$7\% of all CG flashes) are well correlated with annual values
  of CONUS CP.  The relatively strong relations between CP and CG
  flash counts in some regions and times of the year at daily
  resolution provide a benchmark for more complex proxies and suggest
  that proxy-based extended- and long-range prediction of lightning
  activity may be feasible to the extent that precipitation rate and
  CAPE can be predicted.}

\begin{document}

\maketitle

%

\section{Introduction}

Lightning flash rate is a defining characteristic of thunderstorm
evolution. Cloud-to-ground (CG) lightning impacts societies through
deaths and injuries, property damage, wildfires, and air quality
\citep{Koshak2015}.  The importance of CG lightning is a motivation
for studying how its characteristics vary under climate change and
variability.  The relation of lightning with climate, whether in the
form of interannual variability or long-term trends, is difficult to
infer directly from the observational record because high-quality,
spatially-complete lightning datasets are often relatively
short. Moreover, observational data can at best provide circumstantial
information about expected lightning characteristics in climates that
differ from the one in which observations are collected, be they
future climates or ones in other regions.

Lightning activity in general depends on the dynamics and microphysics
of convective clouds, and this dependence has been modeled with
varying levels of detail and complexity.  Lightning occurrence and
flash rates can be simulated with considerable fidelity and realism by
combining electrification and lightning parameterizations with models
of atmospheric dynamics and microphysics
\citep{Mansell2005,Kuhlman2006,Fierro2013}.  Lightning rates have also
been related in a more empirical but effective manner to cloud
properties, microphysical parameters, and updrafts generated by
convection permitting models
\citep{McCaul2009,Yair2010,Lynn2012}. Alternatively, lightning flash
densities can be diagnosed from the output of convective
parameterization schemes in models that do not explicitly resolve
clouds \citep{Allen2002,Lopez2016}.
\citet{Stolz2017} parameterized storm-scale total lightning density
using environmental variables from reanalysis and aerosol data.

However, detailed cloud and microphysical properties are not always
readily available from reanalysis, seasonal forecasts, or climate
change projections, and are themselves uncertain. Therefore, simple
proxies for lightning
that depend on a few, easily available quantities may provide utility
for climate variability and projection applications.
\citet{Romps2014} proposed the product of convective available
potential energy (CAPE) and total precipitation rate as a proxy for
the number of total CG lightning flashes. The application of this
proxy to climate change projections predicts a 50\% increase in United
States lightning strokes over the 21st century.  An attractive feature
of the Romps proxy is that there is some theoretical understanding of
how its constituents are modulated by short-term climate variability
\citep[e.g., ENSO;][]{Ropelewski87,LHeureux:ENSO2014,Allen:ENSO2014}
and long-term change \citep{HeldSoden2006,Seeley2016}.  Also, its
ingredients are standard outputs of many seasonal and subseasonal
dynamical forecasting systems and could be used to make extended-range
forecasts of lightning activity \citep{Dowdy2016,Munoz2016}. Analogous
approaches have been used to relate tornado and hail activity with
nearby meteorological quantities at monthly and daily resolution
\citep{Tippett2012:grl,JAllen:Hail2014,TippettCohen:ExtremeTornado,Westermayer2017}. A
caveat of empirical proxy-based approaches is that good performance in
the current climate does not guarantee good performance in future
climates \citep{Stainforth2007,CamargoHIRAM2013}.

The goal of this work is to assess the extent to which the ingredients
of the Romps proxy, precipitation rate and CAPE, capture recent
variability in CG flash counts over the contiguous United States
(CONUS).  Of course, assessing the performance of the proxy in
capturing observed CG flash counts is a necessary, but not sufficient
requirement for its use in applications such as climate projections
and subseasonal to seasonal (S2S) predictions, which is our long-term
goal. Importantly, good performance with reanalysis in no way
guarantees comparable performance in climate projection or S2S
forecast applications. On the other hand, poor performance of the
proxy in reanalysis would provide useful indications of its
limitations.  Also, the choice of the Romps proxy is motivated by the
availability of CAPE and total precipitation rate in forecast and
reforecast datasets such as those of the NOAA Climate Forecast System,
version 2 \citep{Saha2011,Lepore:CFSv2:2018}, the S2S Prediction
Project Database \citep{Vitart:S2S:BAMS}, and the Subseasonal
Experiment (SubX) database \citep{subx:bams}. However, the CAPE values
that are available from forecast models can be sensitive to the choice
of parcel and use of the virtual temperature correction.

\citet{Romps2014} showed a strong association between daily counts of
total CONUS CG flashes and the product of CONUS-averages of
precipitation rate and CAPE during a singer year, 2011. Here we use a
longer period (14 years) and to see what additional years of data
can tell us about seasonal variations in the strength of the
association and about the ability of the proxy to capture interannual
variability in CG flash counts.  \citet{Romps2014} used CAPE
calculated from radiosonde data and a NOAA River Forecast Centers
precipitation product based on rain-gauge and radar data.  Here we
test the quality of the association between CG flash counts and the
CAPE-precipitation product when reanalysis products are used instead
of solely observation-based ones.  Knowing whether or not reanalysis
products are sufficiently realistic for this purpose is important
because it helps to judge whether such a climate proxy computed from
numerical model outputs can be used to forecast or project future CG
lightning activity.  Furthermore, the use of spatially complete
reanalysis data allow us to form the product of collocated CAPE and
precipitation rate (denoted CP) on a spatially resolved grid and to
examine regional features of the association of the CAPE-precipitation
product with CG lightning flash counts.


\section{Data and methods}
\subsection{Data}
We use precipitation rate (mm d$^{-1}$) and 0-180 hPa most unstable
CAPE (J kg$^{-1}$) data from the North American Regional Reanalysis
\citep[NARR;][]{Mesinger2006:NARR}. The most unstable parcel is found
by dividing the 0-180 hPa layer into six 30-hPa-deep layers and
selecting the one with the largest equivalent potential temperature
(no virtual temperature correction).  NARR data are provided at
3-hourly resolution. The precipitation rate is based on a 3-hour
accumulation. CAPE is the instantaneous value at the start of the
3-hour period.  The data are averaged from their 32 km native grid
spacing to a $1^\circ \times 1^\circ$ latitude-longitude grid.  We
choose the 1-degree grid spacing to match that of CFSv2 and SubX data
because our long-term goal is S2S prediction.  NARR precipitation
estimates show some advantage over ones from other reanalysis
products, especially global reanalysis, likely due to its use of
precipitation observations which it assimilates as latent heating
profiles \citep{Bukovsky2007,Cui2017}.

Cloud-to-ground (CG) lightning flash counts come from the National
Lightning Detection Network \citep[NLDN;][]{CumminsMurphy2009} and are
summed on the same $1^\circ \times 1^\circ$ grid at daily (UTC)
resolution.  Only CONUS land points are used in our analysis although
both NARR and NLDN data extend over ocean and into Mexico and Canada.
We use NLDN data covering the period 2003--2016 (5114 days) during
which the NLDN network is complete and stable 
\citep{Koshak2015}.  There are 9 days with missing NLDN data:
31/12/03, 08/02/04, 08/02/06, 16/01/07, 06/02/07, 18/12/07, 05/12/08,
17/01/09, 16/01/13, and those days are excluded from calculations.  A
change in the NLDN Total Lightning Processor (TLP) on 18/09/15
may be responsible for a substantial increase noted in the number of
positive CG flashes reported during 2016 \citep{Nag:2016update}.

\begin{figure}[h]
\begin{center}
\includegraphics{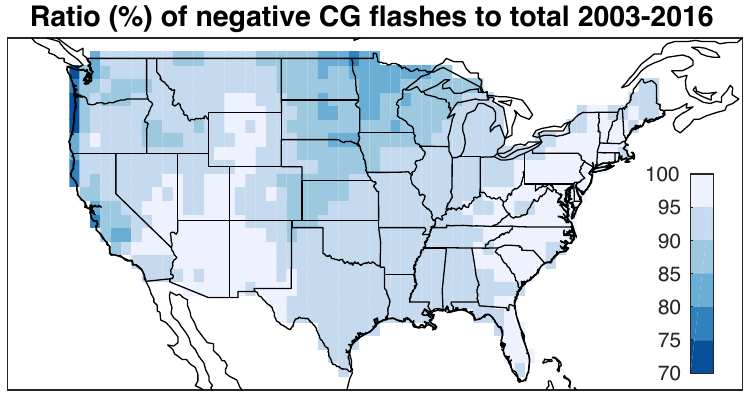}
\end{center}
\caption{Ratio in percent of the number of negative polarity CG
  flashes to the total number of CG flashes 2003--2016.}
\label{fig:ratioNT}
\end{figure}

Negative and positive ($\ge$ 15 kiloampere; kA) polarity CG lightning
flash counts are analyzed separately to examine whether the proxy
performance differs for positive and negative flash counts.  Differing
proxy performance might be expected since environment, at least at the
mesoscale, can affect storm structure and CG flash polarity
\citep{Carey2007}. Total CG flashes were also analyzed but give
results that are very similar to those for negative CG flash counts.
The threshold of 15 kA for positive polarity flashes accounts for the
tendency of the NLDN to misclassify cloud pulses as low-amplitude,
positive CG strokes \citep{Biagi2007,CumminsMurphy2009}.  The new NLDN
TLP removes the 15 kA peak current limit for positive CG
flashes. Therefore, to maintain more consistency, we apply our own 15
kA filter to the positive CG flashes in 2015 and 2016. As a
consequence, the numbers of positive CG flashes analyzed here for 2015
and 2016 are less than those in the unfiltered NLDN data sets.  The
majority (93\%) of CONUS CG flashes have negative polarity during the
period 2003--2016.  The ratio of negative polarity CG flashes to total
CG flashes shows substantial spatial variations (Fig.\
\ref{fig:ratioNT}), with ratios above 95\% on the Eastern Seaboard,
below 85\% in the Upper Midwest and the lowest values in very narrow
band along the West Coast, consistent with the patterns found in
individual years \citet{Koshak2015}.  Positive polarity CG flashes
have been associated with severe thunderstorms in the Midwest
\citep[tornadoes, hail, and damaging wind;
e.g.,][]{MacGorman1994,Carey2003}.  However, other regions with
frequent severe weather do not show especially elevated percentages of
positive CG flashes. Conversely, the high percentages of positive CG
flashes along the West Coast occur where severe weather is infrequent
\citep{Zajac2001,Koshak2015,Medici2017}.

Here we use the product of {collocated} 3-hourly values of CAPE
and precipitation rate as a proxy for the number of CG flashes that
occur during those three hours in the corresponding grid cells, and we
denote this quantity as CP.
Using a proxy based on collocated values allows us to compare it with
CG flash counts on a regional as well as CONUS-wide basis.
Three-hourly CP values are summed over time to form daily, monthly, seasonal,
and annual values; they are summed over space to form regional CONUS values.
CP values are scaled to facilitate their comparison with CG flash
counts. The scaling factor is computed so that the area-weighted sum
of CP values over the period 2003--2016 matches the number of CONUS
flashes, depending on polarity. In other words,
\begin{equation*}
  \text{scaling factor} = \frac{\text{area-weighted sum (CP)}}{\text{sum (CG
    flashes)}}\,.
\end{equation*}
The scaling factor for negative polarity CG flashes is 64.17 flashes /
J kg$^{-1}$ mm day$^{-1}$ and 4.79 flashes / J kg$^{-1}$ mm day$^{-1}$
for positive polarity CG flashes. On the
$1^\circ \times 1^\circ$ grid,
\begin{equation*}
\text{CP (scaled to negative CG flash counts)} = 64.17 
\times \text{CAPE}
\times \text{precipitation} \times \cos \phi\,,
\end{equation*}
and
\begin{equation*}
\text{CP (scaled to positive CG flash counts)} = 4.79 
\times \text{CAPE}
\times \text{precipitation} \times \cos \phi\,,
\end{equation*}
where $\phi$ is latitude in radians, and $\cos \phi$ accounts for the
varying grid cell area.
The same scaling factor is used in all months and locations. All
comparisons between CP and CG flashes use scaled CP, and we drop the
word scaled hereafter.

\subsection{Methods}
We assess regional behavior by spatially aggregating CP and CG flashes
at the level of NOAA climate regions \citep{Karl1984}. The states in
each region are listed in Table \ref{tab:regions}. The spatial
structure of CP and CG flashes are compared using pattern correlations
computed for the points east of 105$^\circ$W with the map mean removed
\citep{Wilks2011}. The pattern correlations are computed using points
east of 105$^\circ$W to focus on the region where the vast majority of
CG flashes are recorded and to avoid giving credit for simply matching
the east-west gradient.  The temporal association between CP and CG
flashes is measured using correlation and mean-squared error (MSE),
where the error is the difference of CP and the number of CG flashes.
MSE is normalized by the CG flash variance (at the same temporal
resolution) to allow comparison of error levels for regions and
seasons with disparate levels of CG flash activity.  In addition to
daily, monthly, and annual aggregation, we also look at totals for
six-month warm (May--October) and cool (November--April) seasons, with
the cool season comprising the 13 complete seasons
2003/2004--2015/2016 for which data are available. For interannual
correlations (14 years), the critical correlation value at the 5\%
significance level for rejecting the null hypothesis of no correlation
is about 0.46 for a one-tailed test and 0.53 for a two-tailed test.

\section{Results}
\subsection{Regional scaling of CP and CG flashes}
Although CP is scaled to match the CONUS total of CG flashes, the
ratio of CG flashes to CP shows distinct regional variations (Fig.\
\ref{fig:ratio}), which presumably reflect the differing frequency of
rainfall processes and cloud properties that are not accounted for in
CP.  Also, the spatial variations in the ratio of CG flashes to CP may
indicate a role for additional thermodynamic factors such as wet-bulb
temperature \citep{Koshak2015}, mid-level humidity
\citep{Westermayer2017} or warm cloud depth \citep{Stolz2017}. Values
of the ratio of negative CG flashes to CP are slightly above one in
the Northeast, where aerosol concentrations are relatively large
\citep{PM2.5}.  The ratio of negative flashes to CP is near one east
of the Rockies, substantially greater than one west of 105$^\circ$W,
until along the West Coast where the ratio is much less than one. The
picture for positive CG flashes is similar, but without elevated ratio
values in the Northeast and with a stronger gradient from values
greater than one in the Northern Rockies and Upper Midwest to values
less than one in the Southeast.  The single scaling factor applied to
CP for each polarity matches the behavior in the East because 91.0\%
of the negative polarity CG flashes and 93.4\% of the positive
polarity CG flashes occur east of 105$^\circ$W. Previous studies have
noted that ice-based precipitation processes are dominant in the arid
Southwestern US, and the ratio of convective rainfall to lightning
(rainfall yield) is relatively low there \citep{Petersen1998}.
\citet{Mulmenstadt2015} found that ice-phase clouds were more frequent
in the western half of the CONUS and that liquid-phase clouds were
more frequent east of the Rockies. Lower lifted condensation level
heights in the East (not shown) are suggestive of lower cloud bases
and conditions favoring warm rain processes with greater precipitation
efficiency.  \citet{Fuchs2015} compared total lightning flash rates in
Colorado, Oklahoma, Alabama, and the District of Columbia, and
hypothesized that storms with high cloud base heights or shallow warm
cloud depths have less warm-phase precipitation and more mixed-phase
precipitation and lightning.  The low ratio of CG flash count to CP in
a narrow band along the West Coast may be related to onshore flow of
maritime air masses with fewer cloud condensation nuclei or
dynamically weaker convection that develops offshore and at coastal
boundaries \citep{Zipzer1994,Xu2013,Holle2016}.  Despite some regional
scaling deficiencies, CP matches CG flash counts relatively well in
the areas where the vast majority of CONUS CG lightning occurs.

\subsection{Daily associations of CONUS totals}\label{sec:daily}
We first consider the association of daily CONUS CG flash counts with
CAPE alone.  The correlation of daily CONUS CG flash counts with
CONUS-averaged CAPE from NARR over the period 2003--2016 is larger
(values of 0.86 and 0.8, respectively, for counts of negative and
positive CG flashes; Fig.\ \ref{fig:scatter}) than that found by
\citet{Romps2014} for CAPE computed from radiosonde data for the
single year 2011 ($r=0.72$).  In fact, there is some expectation that
reanalysis CAPE might be more representative of large-scale features
than would be CAPE computed from radiosonde data, since radiosonde
data contain small-scale variability that may not be representative of
the large-scale nearby environment.  \citet{Lepore2015} found a
stronger relation between gauge-measured rainfall extremes and CAPE
from reanalysis than with CAPE based on nearby radiosonde
measurements.  Also, the use of time-averaged reanalysis CAPE and
precipitation possibly mitigates the difficulty of using observed
collocated CAPE and precipitation that is caused by CAPE being
released by convection \citep{Romps2014}.  On the other hand, daily
CONUS-averaged precipitation from NARR shows a slightly weaker
relation with CG flash counts (correlation values of 0.36 and 0.43 for
negative and positive CG flash counts, respectively; Fig.\
\ref{fig:scatter}) than the $r=0.54$ reported by \citet{Romps2014}
using a NOAA River Forecast Centers precipitation product based on
rain-gauge and radar data.  Daily CONUS CP has a slightly stronger
relation with flash counts (correlation values of 0.89 and 0.87 for
negative and positive polarity, respectively; Fig.\ \ref{fig:scatter})
than does CAPE alone. We note that including precipitation rate has
the potential to introduce some dependence on aerosols since rain rate
increases with aerosol level \citep{Koren2012}.

\subsection{Annual cycle and climatology}
CONUS CP and CG flash counts have strikingly similar annual cycles,
both at daily and monthly resolution, with peak values in summer and
much smaller values in the cool season (Fig.\ \ref{fig:annual}).  The
increase in CG occurrence from spring to summer is gradual and
followed by a somewhat sharper decay after August \citep{Holle2016}.
Annual cycles at daily resolution are computed by averaging the 14
values (2003--2016) available for each calendar day with February 29
excluded.  The daily resolution annual cycle shows that CONUS CP
appears to resolve some sub-monthly features that are likely specific
to this set of years.  The largest discrepancy between the annual
cycles of CONUS CP and negative CG flash counts (clearest at monthly
resolution) occurs in July--August when CP values are too low and in
September--October when CP values are too high (Fig.\
\ref{fig:annual}).  There is also good agreement between the annual
cycles of CONUS CP and positive CG flash counts, with a tendency of CP
values to be too low in spring (March--May) and too high in summer and
early fall (June--September).

The similarity of the annual cycles of CONUS CP and CG flash counts
means that some of the variance of CG flash counts at daily resolution
explained by CP (Fig.\ \ref{fig:scatter}) is a consequence of CP
accurately capturing the seasonality of CG flash counts. In fact, the
annual cycle of CP at daily resolution, a quantity with no
year-to-year variation, explains 63\% and 52\% of the variance of
negative and positive daily CONUS CG flash counts, respectively.  The
annual cycle of CP at daily resolution explains nearly as much
variance of daily negative and positive CONUS CG flash counts as do
their own annual cycles, which explain 66\% and 54\%, respectively.

CP captures the annual cycle of lightning occurrence at the regional
level and monthly resolution to varying degrees (Fig.\
\ref{fig:regional_clim}).  Because a single (polarity-dependent)
factor is used to scale CP, the differing performance of CP in
matching the relative magnitude and phasing of the regional seasonal
cycles provides another indication of the extent to which CAPE and
precipitation alone are adequate to provide a statistical description
of CG flash counts.  For the most part, both the magnitude of the CG
annual cycle and its phasing are well-matched by that of CP in regions
east of the Rockies. Because the majority of lightning flashes occur
in the eastern half of the country, the scaling factor is disposed to
match the magnitude there.  In the Upper Midwest, where the ratio of
positive to negative CG flashes is relatively large, the CP annual
cycle is stronger than that of the negative CG flashes, and better
matches the annual cycle magnitude of the positive CG flashes.  CP
overestimates negative CG flashes in the Upper Midwest during
July--August and underestimates them in the Plains.  The largest
differences in annual cycle magnitude are present in the Southwest,
Northwest, and West regions where CP is substantially too low compared
to both negative and positive CG flash counts. These annual cycle
biases indicate that a lightning proxy could potentially benefit from
taking into account physical factors that are different in these
regions (e.g., warm cloud depth) and that are not captured by
reanalysis precipitation and CAPE.  Alternatively, regionally-varying,
empirical corrections could be applied to CP as is done to the output
of numerical weather and climate prediction models.

CP shows the largest phase errors in the Northwest and West. CP in the
Northwest peaks in May--June while CG flash counts peak in June--August
and have stronger seasonality (greater peak to trough differences).
In the West, CP shows a bimodal structure with a peak in August and a
secondary peak in early spring, while CG flash counts have a unimodal
distribution with a peak in July and stronger seasonality.  CP tends
to match the annual cycle of positive CG flashes less well than it
does negative ones, especially in the Southeast, Northeast and Central
regions where CP overestimates the peak magnitude. 

\begin{figure}[h]
  \centering
\includegraphics{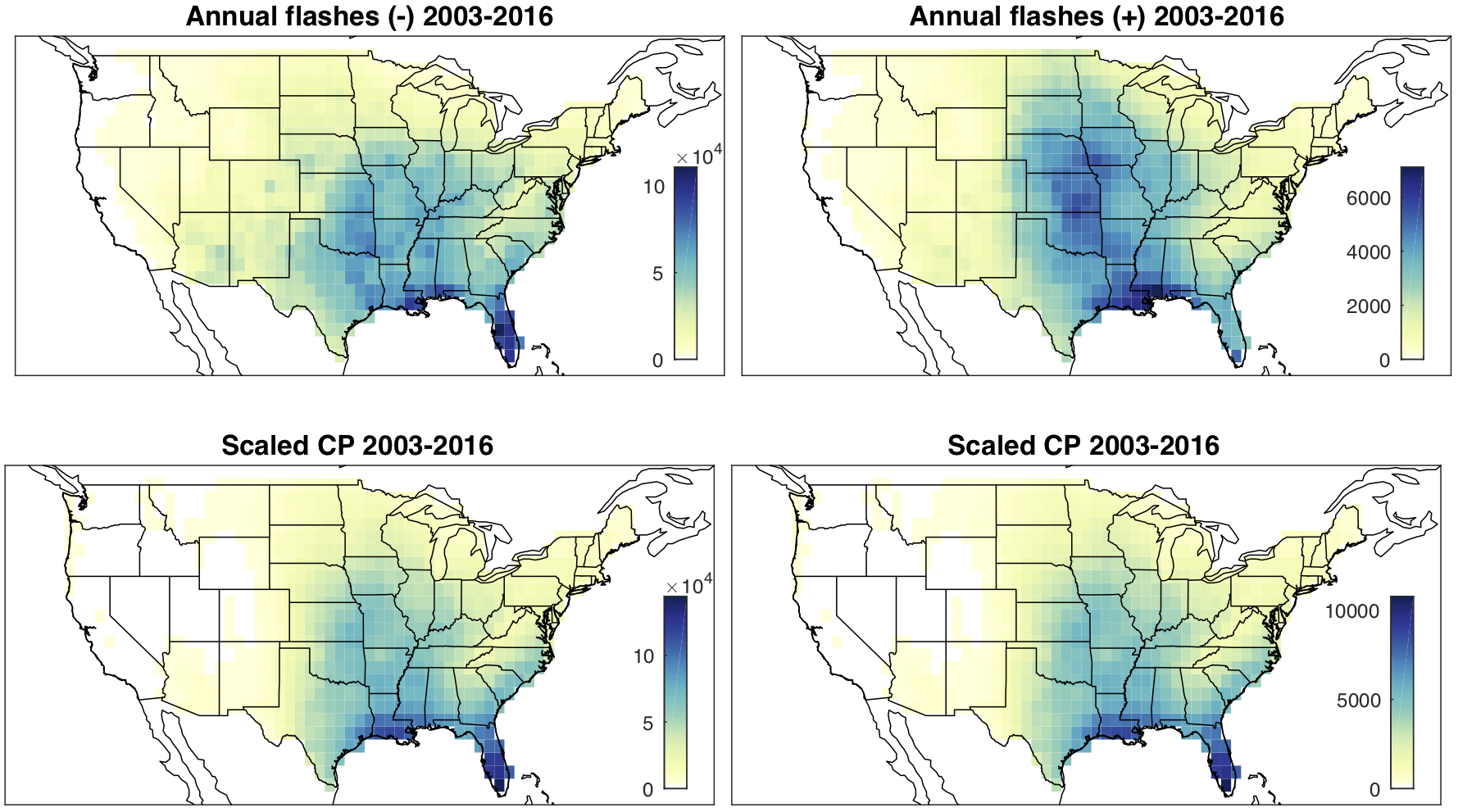}
\caption{Annual averages of negative and positive polarity lightning
  flashes and the corresponding maps of CP.}
\label{fig:climmap}
\end{figure}

The annual spatial distribution of CP is more similar to that of
negative CG flashes than that of positive CG flashes (Fig.\
\ref{fig:climmap}).  The annual spatial distribution of negative CG
flashes is nearly indistinguishable from all CG flashes (not shown).
Compared to counts of positive CG flashes, corresponding CP values are
too low in the middle of the CONUS (Oklahoma, Kansas, and Nebraska) and
too large along the Southeast coast, and over Florida.  Centered
pattern correlations between the climatological monthly maps of NLDN
flash counts and the CP proxy computed for the CONUS region east of
105$^\circ$W show good agreement for negative polarity flashes
throughout the year, but are relatively poor for positive CG flashes
during June--October (Table \ref{tab:pc}), which are months when CP
values tend to be too high in the Southeast, Northeast, and Central
regions and too low in the Plains region (Fig.\
\ref{fig:regional_clim}).

\subsection{Seasonality in the strength of daily associations}
We compute the correlation between daily CONUS CG flash counts and CP
values for each of the twelve calendar months separately to examine
how the strength of the association between daily CP values and flash
counts varies through the calendar.  By removing the mean of each
calendar month from the daily data, we remove much of the contribution
of the annual cycle to the correlation of daily values computed in
Section 3\ref{sec:daily}, though some months have considerable
climatological mean changes within the month (e.g., August). The
correlation between daily CONUS CP and CG flash counts by calendar
month is very similar for negative, positive, and all polarity flashes
(Table \ref{tab:dailycor}), and shows a clear seasonality with lower
values in summer and fall.  The median correlation between daily CONUS
CP and all CG flash counts by calendar month is 0.88 during the months
of November through May, while it is 0.70 during the months of June
through October.

The normalized daily MSE shows a consistent picture with larger
relative errors in the months of June through November (Table
\ref{tab:dailycor}).  The daily normalized MSE is relatively low
during the months of November through April with a median of 24\% and
36\% for negative and positive CG flash counts, respectively. However,
during the months of May through October, the median daily normalized
MSE is 66\% and 65\%, respectively, for negative and positive CG flash
counts.  Normalized MSE values greater than one for negative CG
flashes in September and October indicate that the errors are greater
than would result from replacing the daily CP values with the monthly
average over the period.  Higher daily errors during the months of
peak lightning occurrence might be expected to accumulate and limit
the ability of CP to capture year-to-year variations of seasonal and
annual totals.

\subsection{Interannual variability}
We now examine the ability of CP to match year-to-year variations in
monthly, six-month, and annual values of CG flash counts.
At annual resolution, there is no substantial correlation of CONUS CP
with counts of negative CG flashes ($r=0.09$; Fig.\
\ref{fig:annual.CP}) or with counts of all CG flashes ($r=0.03$, not
shown). To some extent, this lack of association between annual values
of CONUS CP and negative CG flashes is unexpected since daily values
are well-correlated, both in an overall sense that includes
seasonality, and when focusing on individual months.  Time averaging
often reduces noise and enhances statistical relations, but, in this
case, time averaging serves to highlight deficiencies of CP during the
warm season.
%
%
The correlation of annual CONUS CP with counts of positive CG flashes
is considerably larger ($r=0.8$; Fig.\ \ref{fig:annual.CP}), although
that correlation drops to 0.52 when the largest annual value (2016) is
removed.  The large number of positive CG flashes recorded in 2016
could be due to the new flash type classification technique that is
based on the examination of multiple waveform parameters, and that was
employed within the TLP as indicated in \citet{Nag:2016update}.


The correlation between warm season (May--October) CONUS CP and CG
flashes is roughly the same as for annual values ($r=0.05$ and
$r=0.76$ for negative and positive polarity, respectively; Fig.\
\ref{fig:annual.CP}).  Annual counts of CG flashes are dominated by
warm season values in the sense that warm season flashes account for
89\% of negative CG flashes and 83\% of positive CG flashes during the
period 2003--2016 (Tables \ref{tab:percent.neg} and
\ref{tab:percent.pos}), and in the sense that warm season counts of CG
flashes are highly correlated with annual values ($r=0.97$ and
$r=0.99$, for negative and positive flashes, respectively).  During
the cool season (November--April), CONUS CP shows a fairly good
association with negative ($r=0.74$; Fig.\ \ref{fig:annual.CP}),
positive ($r=0.89$; Fig.\ \ref{fig:annual.CP}) and all ($r=0.79$, not
shown) CG flash counts.
The good association of CONUS CP with the number of CG flashes of both
polarities during the cool season is possible because the correlation
between the number of negative and positive cool season CONUS CG
flashes is 0.69. In contrast, the correlation between the number of
negative and positive warm season CONUS CG flashes is -0.32.

The correlations of monthly values of CONUS CP with negative CG
flashes (first line of Table \ref{tab:neg.corr}) are considerably
higher than the annual or warm season correlations, though the
correlations for months during the warm season (median correlation
0.47) tend to be lower than for months during the cool season (median
correlation 0.87).  Correlations of positive CG flash counts with
monthly CONUS CP are above 0.85 during six months of the year and fall
below 0.5 only in August and October (first line of Table
\ref{tab:pos.corr}), with a tendency toward lower correlations during
warm season months (median correlation 0.65) compared to cool season
months (median correlation 0.87).  The lower level of correlations
during some warm season months is consistent with relatively low
correlations and poor normalized MSE at daily resolution during warm
season months noted earlier (Table \ref{tab:dailycor}).


At the regional level, correlations between monthly CP and negative CG
flash counts are overall higher than those at the CONUS level,
particularly during warm season months (May through October) when 87\%
of the regional correlations are greater than CONUS ones (Table
\ref{tab:neg.corr}).  Lower correlation with increasing spatial
aggregation is consistent with regionally varying magnitude errors.
Correlations of monthly CP and negative CG flash counts show some
indication of relatively lower values in warm season months in the
Southeast, Northeast, and South regions, which produce more than 68\%
of the annual CONUS number of negative CG flashes (Table
\ref{tab:percent.neg}).
Correlations of monthly CP and positive CG flash counts
show some reduced values during warm season months in the South and
Southeast regions (Table \ref{tab:pos.corr}), but are otherwise fairly
strong except in the Central, Upper Midwest, and Plains regions during
some cool season months.  Despite clear deficiencies in the South, and
Southeast regions during the warm season, which likely contribute to
poor CONUS performance, the relation between monthly CP and CG flash
counts is strong in many regions and during many times of the year,
demonstrating the strong potential for CP as an indicator of monthly
tendencies in regional CG flash counts.

Maps of the correlation between warm and cool season CP and CG flashes
show large-scale features that are consistent with the analysis at the
NOAA region level, as well as fairly high correlations at the
gridpoint level in many areas (Fig.\ \ref{fig:map.cor}).  Correlations
of warm season CP with negative CG flash counts are mostly positive,
but very modest in the Southeast, Northeast, and parts of the South,
consistent with the regional analysis.  Correlations between warm
season CP and negative CG flash counts are weakly negative in an area
that includes the borders of Colorado, Wyoming, Nebraska, and Kansas
where the ratio of intra-cloud lightning to CG flashes is known to be
large and which is often associated with inverted or complex charge
structures \citep{Carey1998,Medici2017}. Misclassification of cloud
pulses as CG flashes as well as incorrect assignment of first peak
polarity have been noted in the Kansas-Nebraska area
\citep{CumminsMurphy2009}.  Warm season correlations between CP and
negative CG flash counts are also weakly negative in smaller areas
along the West Coast, and around Butte, Montana, and Columbus,
Georgia. Cool season correlations between CP and CG flash counts of
both polarities are similar and generally higher than warm season ones
across the southeastern half of the CONUS.  Correlations between warm
season CP and positive CG flashes are overall higher than for negative
CG flashes over most of the CONUS, with low correlations mostly
limited to Florida and states to its immediate north (Fig.\
\ref{fig:map.cor}).  Maps of annual correlations (not shown) are
similar to those for the warm season.



Warm season normalized MSE exceeds one in many areas for both
polarities, indicating errors that are larger than the climatological
variance, especially west of 105$^\circ$W where mean biases are large
(Fig.\ \ref{fig:map.norm.mse}).  For negative CG flashes, the warm
season normalized MSE east of the Rockies is mostly less than one
except in areas that include Wisconsin and eastern Minnesota, the Gulf
Coast, and eastern North Carolina. On the other hand, the warm season
normalized MSE for positive CG flashes is large across the Southeast
and Northeast regions, indicating magnitude miscalibration since
correlations are good there (Fig.\ \ref{fig:map.cor}).
The normalized cool season MSE is lower overall than the warm season MSE for both polarities,
except in the Northwest and on the West Coast.  Cool season normalized
MSE is lower overall for negative CG flashes than for positive
ones. Both polarities have large normalized MSE over southern Florida
in the cool season. Annual normalized MSE maps (not shown) are similar
to warm season ones.

\section{Summary and discussion}
We have compared the product of collocated CAPE and precipitation
(denoted CP) taken from the North American Regional Reanalysis with
counts of negative and positive cloud-to-ground (CG) lightning flashes
from the National Lightning Detection Network (NLDN) over the
contiguous U.S. (CONUS). Our analysis includes CONUS-wide and regional
characteristics on daily, monthly, and annual resolution for the
period 2003--2016. This analysis extends the findings of
\citet{Romps2014} who considered one year of CONUS-aggregated daily
values from 2011. Overall, the association of CP with lightning
flashes on the daily, monthly, and seasonal scale tends to be stronger
during the cool season (November--April) than during the warm season
(May--October). Interannual correlations between CP and flash counts
tend to be stronger for positive CG flashes than for negative ones.

\begin{figure}[h]
\centering\includegraphics{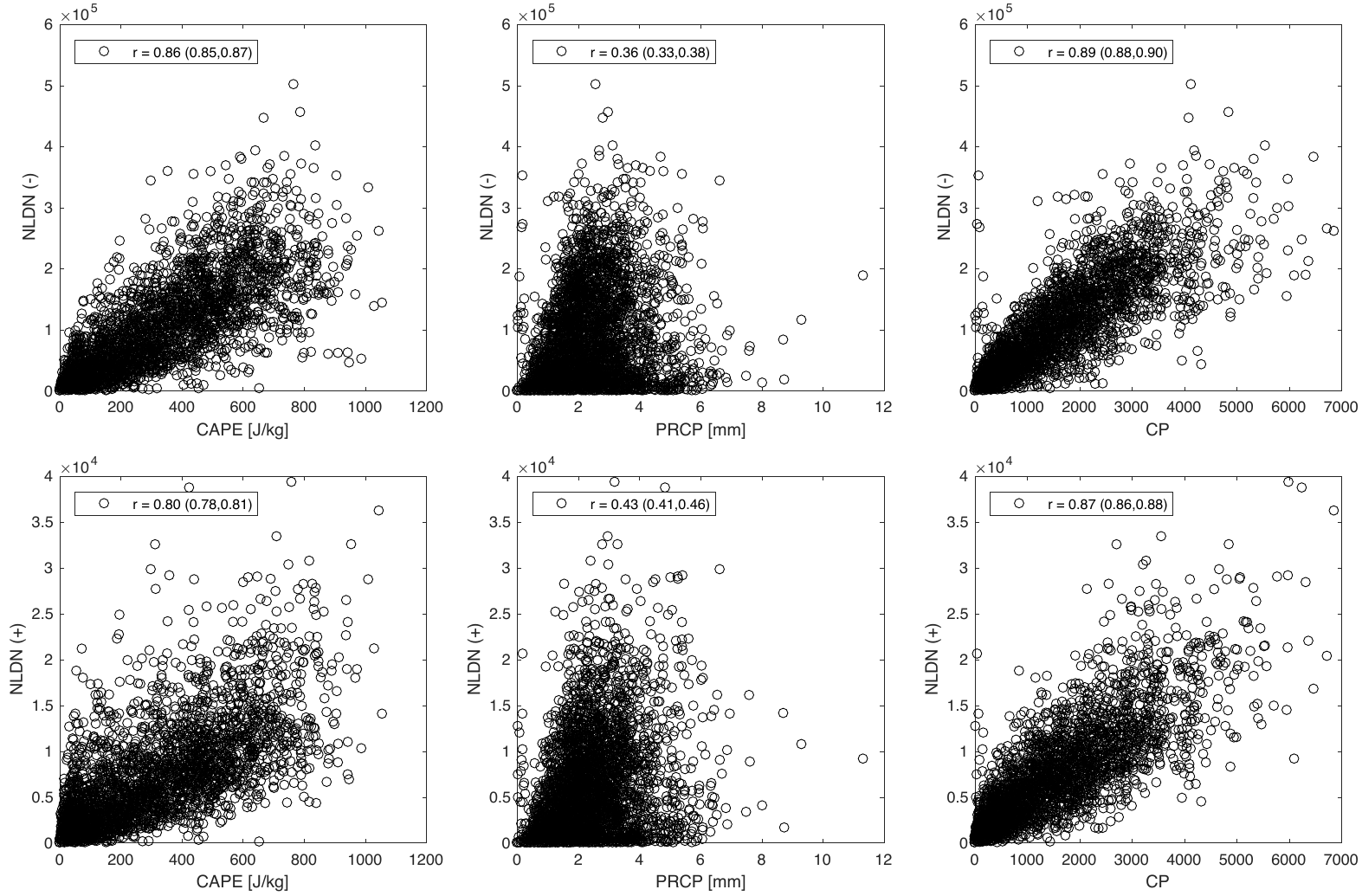}
\caption{Scatter plots of daily values of the CONUS-average of CAPE
  (first column), precipitation (second column), and CP (third column)
  with daily counts of negative (first row) and positive (second row)
  CG flashes. Correlation values and 95\% bootstrap confidence
  intervals (10,000 bootstrap samples) are shown in the legend.}
\label{fig:scatter}
\end{figure}

\begin{figure}[h]
\centering\includegraphics{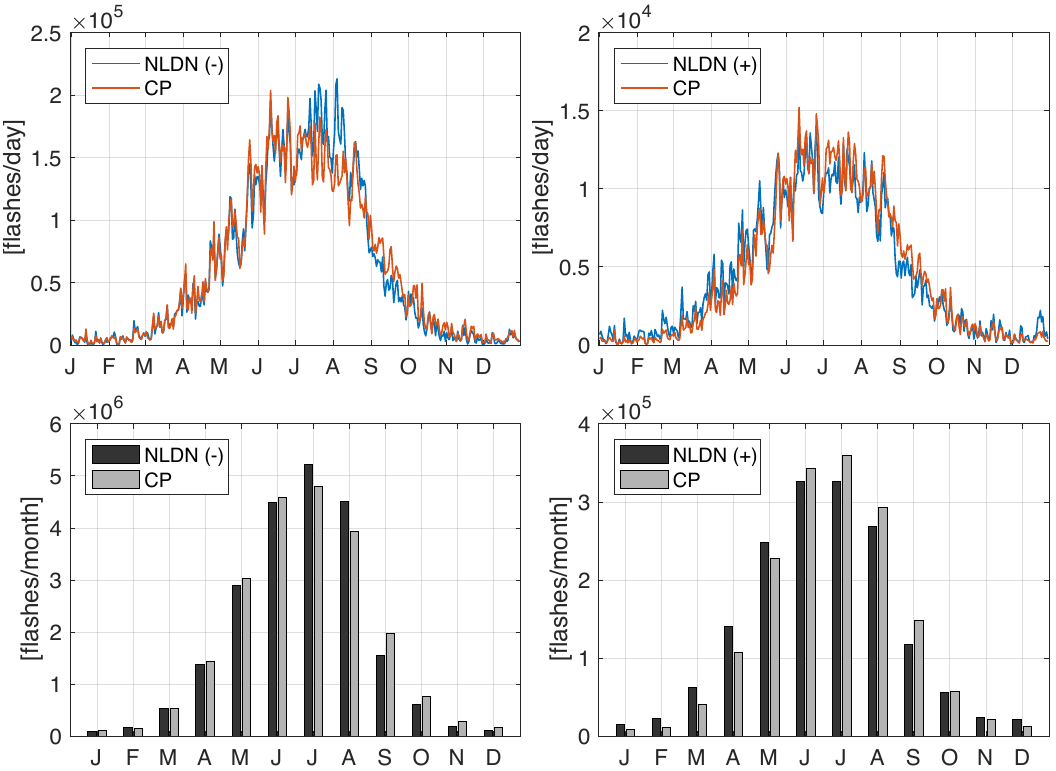}
\caption{Annual cycles of CP and NLDN CG flash counts at daily (top row) and
  monthly (bottom row) resolution for negative (left column) and
  positive (right column) polarity.}
\label{fig:annual}
\end{figure}

\begin{figure}[h]
\centering\includegraphics{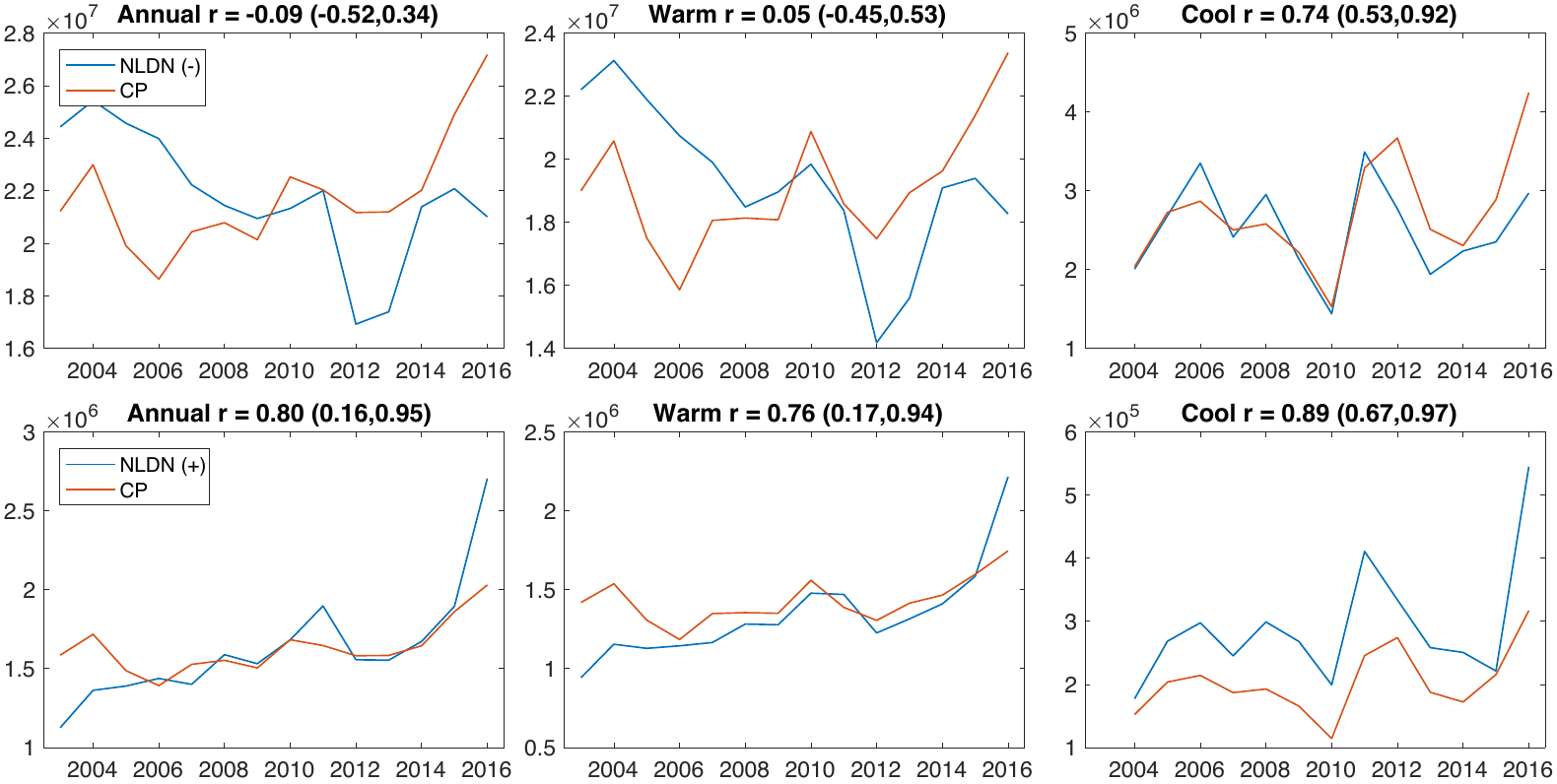}
\caption{Time series 2003--2016 of annual (first column), warm season
  (May-October; second column) and cool season (November--April; third
  column) values of CP and NLDN CG flash counts for negative polarity
  (first row) and positive polarity (second row). Correlation values
  and 95\% bootstrap confidence intervals (10000 bootstrap samples)
  are shown in the titles.}
\label{fig:annual.CP}
\end{figure}

Daily values of CONUS CP are highly correlated with both positive and
negative CG flash counts, explaining more than 75\% of their daily
variance (Fig.\ \ref{fig:scatter}).  Some of this association (more
than 60\% of the variance) is a reflection of the strong seasonal
cycles of CONUS CP and CG flash counts, and their good phase agreement
(Fig.\ \ref{fig:annual}). However, daily variations of CONUS CP and CG
flash counts with respect to their monthly climatologies are still
strongly related, but with stronger associations in the cool season
(November--April) than in the warm season (May--October) when most
lightning occurs (Table \ref{tab:dailycor}). The normalized (relative
to climatological variance) daily mean-squared error (MSE) of CONUS
values is also larger in the warm season, and this lower accuracy
translates into lower interannual correlations
(Fig. \ref{fig:annual.CP}) and higher normalized MSE for monthly
totals in the warm season (Table \ref{tab:neg.corr}), especially for
negative polarity flashes, which are the vast majority of CG flashes.
The low correlation in summer months might indicate that CP is
better-related to storm occurrence than to the total number of CG
flashes within a storm.  This lack of association during the warm
season when the majority of lightning occurs results in there being
essentially no correlation between the annual values of CP and
corresponding numbers of either negative or total CG lightning
flashes.  There is, however, a good correlation between warm season
and annual counts of positive CG flashes with CP
(Fig. \ref{fig:annual.CP}).  Cool season CP correlates well with both
counts of positive and negative CG flashes.

\begin{figure}[h]
\centering\includegraphics{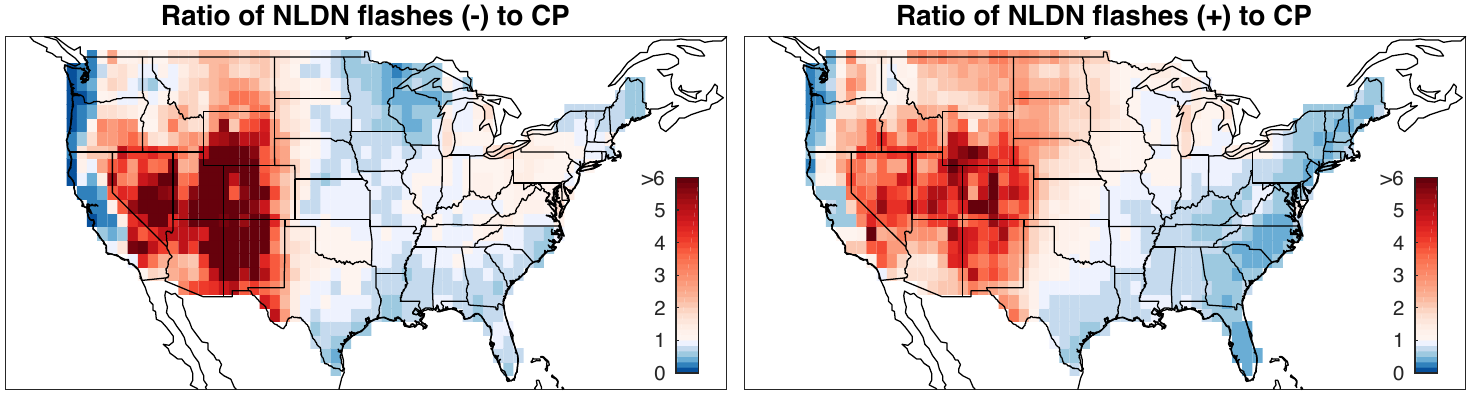}
\caption{Ratio of the number of negative polarity (left) and positive
  polarity (right) NLDN flash counts to CP.}
\label{fig:ratio}
\end{figure}

\begin{figure}[h]
\centering\includegraphics{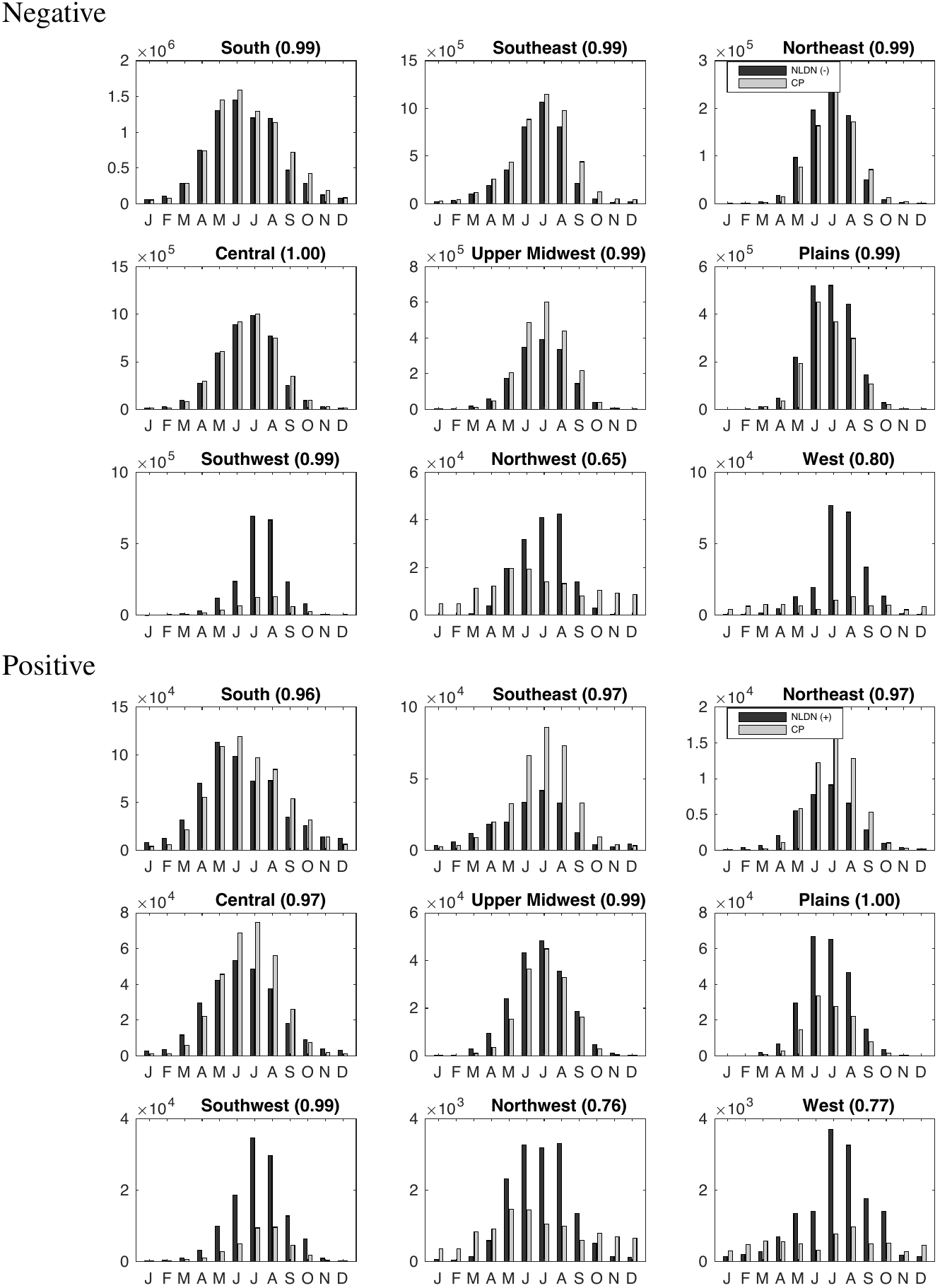}
\caption{Regional climatology for NOAA climate regions. Upper panels
  are for negative CG flash counts and lower panels are for positive
  CG flash counts. The correlation is given in parentheses.}
\label{fig:regional_clim}
\end{figure}

We find that the ratio of CP to CG flash counts varies considerably on
a regional basis, with the greatest difference found in the arid
Southwest where the ratio of CG flashes to CP is substantially higher
than in between regions east of the Rockies (Fig.\
\ref{fig:ratio}). The ratio of CG flashes to CP is lowest along the
West Coast where there are fewer cloud condensation nuclei and where
maritime air masses can penetrate.  At the level of NOAA climate
regions, despite some errors in magnitude, CP matches the annual cycle
in all regions fairly well except the Northwest and West (Fig.\
\ref{fig:regional_clim}).  Correlations of annual CP with CG flash
counts at the regional level are generally higher for positive than
for negative CG flash counts ((Tables \ref{tab:neg.corr} and
\ref{tab:neg.corr}). Regional correlations of annual CP and negative
CG flash counts are especially low in the South and Northeast, where
more than 35\% of all negative CONUS CG flashes occur.  The relatively
weak association between CP and counts of negative CG flashes in these
areas during the warm season explains the lack of correlation between
annual CONUS CP and negative CG flash counts.  In these regions,
correlations between cool season CP and CG flash counts are generally
stronger than for warm season values.

\begin{figure}[h]
\centering\includegraphics{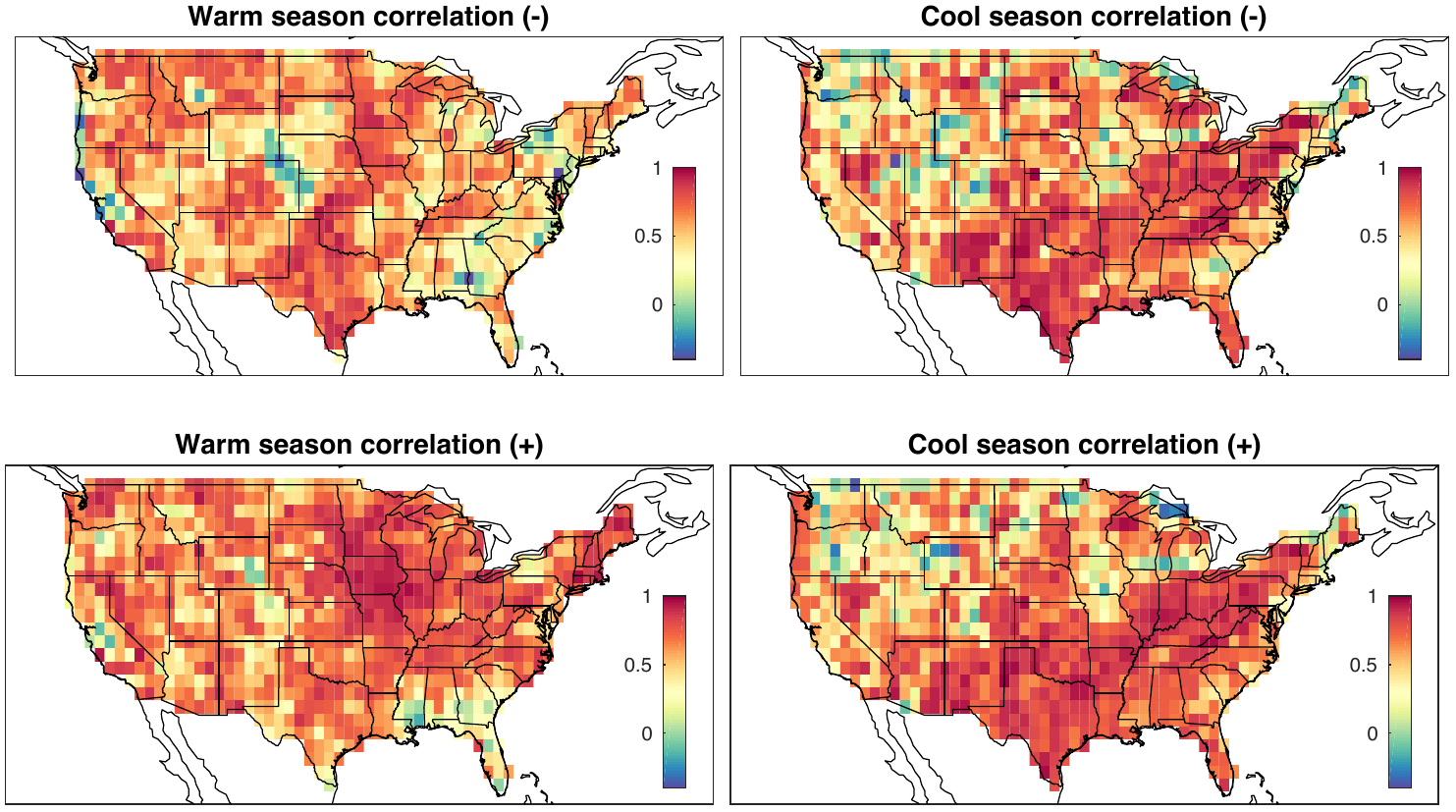}
\caption{Correlation of CP with negative (top row) and positive
  (bottom row) CG flashes during the warm (May--October; left column)
  and cool (November--April; right column) seasons 2003--2016.}
\label{fig:map.cor}
\end{figure}

\begin{figure}[h]
\centering\includegraphics{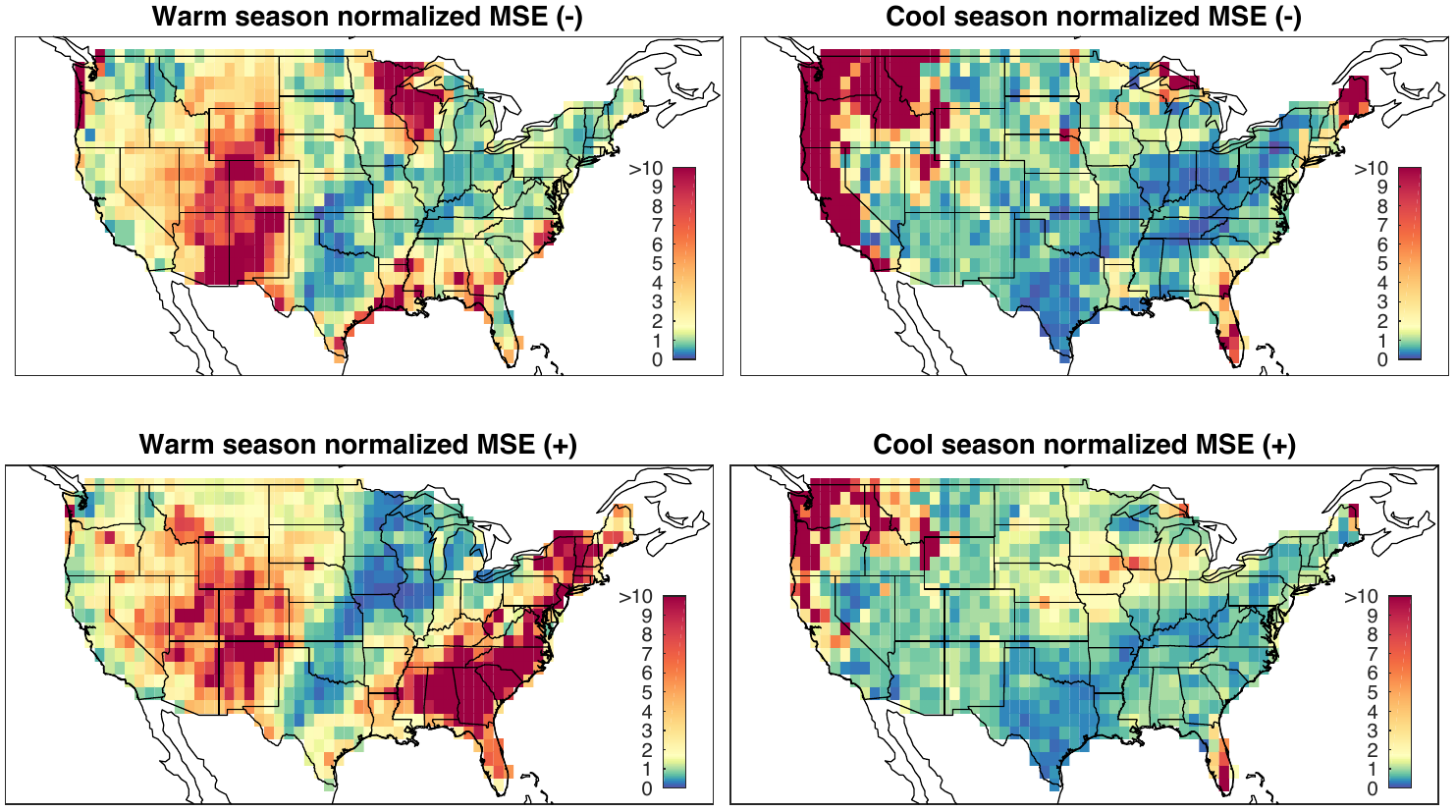}
\caption{Normalized (relative to climatological variance) MSE of the
  difference of CP with negative (top row) and positive (bottom row)
  CG flashes during the warm (May--October; top row) and cool
  (November--April; bottom row) seasons 2003--2016.}
\label{fig:map.norm.mse}
\end{figure}

Maps of the correlation between annual values of CP and CG flash
counts show positive values over most of the CONUS with some spatially
limited exceptions (Fig.\ \ref{fig:map.cor}). In general, correlations
with CP are stronger for positive CG flash counts and stronger during
the cool season for both polarities.  Despite the positive
correlations at the gridpoint level, interannual variability of CP
values are not well-calibrated with CG flash counts.  Maps of
normalized MSE show relatively low error levels for both polarities
during the cool season for most of the CONUS except for large errors
on the West Coast, the Northwest, and southern Florida (Fig.\
\ref{fig:map.norm.mse}). Normalized MSE is small during the warm season
for positive CG flash counts in a swath that extends from Texas
northeast to the Great Lakes.

The suitability of CP for S2S forecasting cannot be concluded from
this study. Further assessment with reforecast data is required
\citep{Tippett:GEFS:2018}. However, in principle, the utility of CP
for S2S forecasting is at least limited by its performance with
reanalysis data and the extent that its constituents can be forecast.
The high correlations between CP and CG flash counts across large
portions of the U.S. show that this simple proxy captures considerable
variability. Conversely, we have also identified regions and times of
the year when CP values are not strongly correlated with CG flash
counts, and this weakness suggests the need for improvements in the
proxy.  The degree to which CP can be predicted in advance is unknown
but there are indications that its ingredients, precipitation and
CAPE, can be predicted with some skill.  U.S. precipitation is already
forecast with current forecast systems with some skill at subseasonal
\citep{DelSoleWeek34:2016} and seasonal \citep{Becker2014}
time-scales. However, skill tends to be lowest in the warm
season. Seasonal values of CAPE have been demonstrated to be
predictable as well \citep{JungKirtman2016}. On the other hand, even
in many locations where correlations between CP and CG flash counts
are high, the MSE is also relatively high, indicating that the CP
proxy is not calibrated to match CG flash counts. In these cases,
there is the potential to correct CP values on a regional basis to
match CG flash counts.  Alternatively, this lack of calibration can
also be interpreted as indicating that the proxy can be improved with
the addition of other factors that are important for characterizing
lightning activity, though perhaps at the risk of losing its
attractive simplicity.

The 14 years of data used in this study are not adequate to answer
fully the question of whether CP is useful proxy for long-term climate
applications. Over the period of study, the annual number of CONUS CG
flashes varied from a high of $2.7 \times 10^7$ flashes in 2004 to a
low of $1.8 \times 10^7$ flashes in 2012, a range of about 35\% of the
annual average of $2.3 \times 10^7$ flashes. However, this variation
in the annual number CONUS CG flashes was not well-captured by the CP
proxy.  This poor performance in describing interannual variability of
CONUS CG flash counts does not necessarily mean that this approach is
poorly suited to climate change applications but does raise concerns
and questions about the applicability of the proxy to climate change
applications. Moreover, there are strong indications that CP is
missing factors whose variation could be relevant for long-term
lightning activity, at least as measured by CG flash counts. Finally,
future work in this area should employ all lightning flashes
(intra-cloud and CG), which is now possible CONUS-wide due to the
recent launch of the Geostationary Lightning Mapper (GLM) on
GOES-16. GLM will provide uniform, continuous total lightning
observations over the Americas and surrounding ocean areas
\citep{GLM2013}.

%
\acknowledgments

Valuable comments and suggestions from Ken Cummins are gratefully
acknowledged.  M.K.T. and C.L. were partially supported by a Columbia
University Research Initiatives for Science and Engineering (RISE)
award; Office of Naval Research awards N00014-12-1- 0911 and
N00014-16-1-2073; NOAA's Climate Program Office's Modeling, Analysis,
Predictions, and Projections program award NA14OAR4310185; and the
Willis Research Network.


We are also thankful for the support from NASA Program
NNH14ZDA001N-INCA (Climate Indicators and Data Products for Future
National Climate Assessments; Dr. Jack Kaye and Dr. Lucia Tsaoussi,
NASA Headquarters).

The authors gratefully acknowledge Vaisala Inc. for providing the NLDN
data used in this study. North American Regional Reanalysis data are
provided by the NOAA/OAR/ESRL PSD, Boulder, Colorado, USA from their
website at http://www.esrl.noaa.gov/psd and the Data Support Section
of the Computational and Information Systems Laboratory at the
National Center for Atmospheric Research (NCAR). NCAR is supported by
grants from the National Science Foundation.

\begin{table}
\begin{center}
\caption{States in NOAA climate regions.}
\label{tab:regions}
\begin{tabular}{lr}
\hline\hline
South &     TX, OK, LA, AR, KS, MS.\\
Southeast &     FL, AL, GA, SC, NC, VA\\
Northeast &     MD, DE, PA, NJ, NY, CT, RI, VT, MA, NH, ME.\\
Central &     TN, MO, IL, IN, OH, KY, WV.\\
Upper Midwest &     IA, MN, WI, MI.\\
Plains &     WY, NE, MT, ND, SD.\\
Southwest &     AZ, NM, UT, CO.\\
Northwest &     OR, ID, WA.\\
West &     CA, NV.\\
\end{tabular}
\end{center}
\end{table}

\begin{table}
\begin{center}
  \caption{Centered pattern correlation between climatological CP and NLDN CG
    flash counts east of 105$^\circ$W.}
\label{tab:pc}
\begin{tabular}{lrrrrrrrrrrrr}
\hline\hline
Polarity & J & F & M & A & M & J & J & A & S & O & N & D \\ 
Total &    0.89 &     0.90 &     0.95 &     0.96 &     0.94 &     0.88 &     0.85 &     0.86 &     0.80 &     0.78 &     0.91 &     0.92 \\ 
Negative &    0.88 &     0.90 &     0.95 &     0.96 &     0.94 &     0.87 &     0.84 &     0.85 &     0.80 &     0.78 &     0.90 &     0.91 \\ 

Positive & 0.89 &     0.87 &     0.92 &     0.95 &     0.88 &     0.68 &     0.51 &     0.67 &     0.69 &     0.75 &     0.90 &     0.91 \\ 
\end{tabular}
\end{center}
\end{table}

\begin{table}
\begin{center}
  \caption{Correlation and normalized MSE of daily CONUS CP and NLDN
    values pooled by month.}
\label{tab:dailycor}
\begin{tabular}{lrrrrrrrrrrrr}
\hline\hline
\multicolumn{13}{c}{Correlation}\\
Polarity & J & F & M & A & M & J & J & A & S & O & N & D \\ 
All &         0.87 &     0.90 &     0.89 &     0.90 &     0.86 &     0.74 &     0.69 &     0.68 &     0.70 &     0.71 &     0.80 &     0.88 \\ 
Negative &        0.87 &     0.90 &     0.89 &     0.89 &     0.85 &     0.73 &     0.67 &     0.66 &     0.69 &     0.70 &     0.79 &     0.87 \\ 
Positive &       0.88 &     0.90 &     0.87 &     0.89 &     0.86 &     0.70 &     0.66 &     0.68 &     0.73 &     0.72 &     0.83 &     0.90 \\ 

\multicolumn{13}{c}{Normalized MSE}\\
Polarity & J & F & M & A & M & J & J & A & S & O & N & D \\ 
All &         0.24 &     0.21 &     0.22 &     0.21 &     0.37 &     0.60 &     0.62 &     0.63 &     1.08 &     0.99 &     0.66 &     0.23 \\ 
Negative &        0.25 &     0.21 &     0.23 &     0.23 &     0.40 &     0.61 &     0.65 &     0.66 &     1.11 &     1.03 &     0.74 &     0.29 \\ 
Positive &        0.40 &     0.39 &     0.33 &     0.27 &     0.29 &     0.67 &     0.64 &     0.64 &     0.77 &     0.66 &     0.32 &     0.42 \\ 
\end{tabular}
\end{center}
\end{table}


\begin{table}
\begin{center}
  \caption{Percent of negative NLDN CG flashes occurring in each NOAA
    region and month 2003--2016. Values of 0.00 indicate less than
    0.01\%.}
\begin{tabular}{lrrrrrrrrrrrrr}
Region & J & F & M & A & M & J & J & A & S & O & N & D & annual\\ 
\hline\hline
South &     0.10 &     0.16 &     0.48 &     0.89 &     1.63 &     3.69 &     4.88 &     3.71 &     0.98 &     0.22 &     0.07 &     0.10 &    16.92 \\ 
Southeast &     0.26 &     0.50 &     1.32 &     3.44 &     5.99 &     6.67 &     5.50 &     5.48 &     2.15 &     1.33 &     0.56 &     0.37 &    33.58 \\ 
Northeast &     0.08 &     0.12 &     0.45 &     1.26 &     2.71 &     4.07 &     4.53 &     3.52 &     1.14 &     0.42 &     0.15 &     0.07 &    18.52 \\ 
Central &     0.01 &     0.01 &     0.08 &     0.27 &     0.79 &     1.60 &     1.80 &     1.53 &     0.66 &     0.19 &     0.04 &     0.00 &     6.98 \\ 
Upper Midwest &     0.00 &     0.00 &     0.05 &     0.22 &     1.01 &     2.39 &     2.39 &     2.03 &     0.66 &     0.13 &     0.01 &     0.00 &     8.89 \\ 
Plains &     0.00 &     0.01 &     0.02 &     0.08 &     0.44 &     0.90 &     1.16 &     0.85 &     0.23 &     0.04 &     0.01 &     0.00 &     3.75 \\ 
Southwest &     0.00 &     0.01 &     0.05 &     0.14 &     0.55 &     1.08 &     3.19 &     3.05 &     1.07 &     0.36 &     0.03 &     0.01 &     9.56 \\ 
Northwest &     0.00 &     0.00 &     0.00 &     0.02 &     0.09 &     0.15 &     0.19 &     0.20 &     0.06 &     0.01 &     0.00 &     0.00 &     0.72 \\ 
West &     0.00 &     0.00 &     0.01 &     0.02 &     0.06 &     0.09 &     0.35 &     0.33 &     0.15 &     0.06 &     0.00 &     0.00 &     1.09 \\ 
CONUS &     0.45 &     0.82 &     2.46 &     6.34 &    13.29 &    20.64 &    23.98 &    20.71 &     7.11 &     2.77 &     0.88 &     0.55 &   100.00 \\ 
\end{tabular}
\label{tab:percent.neg}
\end{center}
\end{table}

\begin{table}
\begin{center}
  \caption{Percent of positive NLDN CG flashes occurring in each NOAA
    region and month 2003--2016. Values of 0.00 indicate less than
    0.01\%.}
\begin{tabular}{lrrrrrrrrrrrrr}
Region & J & F & M & A & M & J & J & A & S & O & N & D & annual\\ 
\hline\hline
South &     0.22 &     0.35 &     0.72 &     1.11 &     1.21 &     2.06 &     2.56 &     2.04 &     0.75 &     0.22 &     0.15 &     0.27 &    11.66 \\ 
Southeast &     0.50 &     0.75 &     1.94 &     4.32 &     6.96 &     6.02 &     4.43 &     4.47 &     2.12 &     1.59 &     0.85 &     0.78 &    34.73 \\ 
Northeast &     0.16 &     0.23 &     0.71 &     1.81 &     2.60 &     3.27 &     2.99 &     2.30 &     1.12 &     0.55 &     0.25 &     0.19 &    16.20 \\ 
Central &     0.02 &     0.02 &     0.19 &     0.58 &     1.48 &     2.65 &     2.96 &     2.18 &     1.15 &     0.30 &     0.08 &     0.01 &    11.62 \\ 
Upper Midwest &     0.00 &     0.01 &     0.13 &     0.41 &     1.82 &     4.09 &     4.00 &     2.86 &     0.93 &     0.21 &     0.02 &     0.01 &    14.50 \\ 
Plains &     0.01 &     0.02 &     0.04 &     0.13 &     0.34 &     0.48 &     0.56 &     0.40 &     0.18 &     0.06 &     0.02 &     0.01 &     2.24 \\ 
Southwest &     0.01 &     0.02 &     0.06 &     0.19 &     0.61 &     1.15 &     2.13 &     1.83 &     0.78 &     0.39 &     0.06 &     0.01 &     7.24 \\ 
Northwest &     0.00 &     0.00 &     0.01 &     0.04 &     0.14 &     0.20 &     0.20 &     0.20 &     0.08 &     0.03 &     0.01 &     0.01 &     0.92 \\ 
West &     0.01 &     0.01 &     0.02 &     0.04 &     0.08 &     0.09 &     0.23 &     0.20 &     0.11 &     0.09 &     0.01 &     0.01 &     0.89 \\ 
CONUS &     0.93 &     1.41 &     3.82 &     8.64 &    15.25 &    20.01 &    20.05 &    16.48 &     7.23 &     3.44 &     1.45 &     1.29 &   100.00 \\ 
\end{tabular}
\label{tab:percent.pos}
\end{center}
\end{table}


\begin{table}
\begin{center}
  \caption{Correlation of monthly averages of negative NLDN flash counts and
    CP. Low-skill values (less than 0.5) are indicated in bold.}
\begin{tabular}{lrrrrrrrrrrrrr}
Region & J & F & M & A & M & J & J & A & S & O & N & D & annual\\ 
\hline\hline
CONUS &     0.85 &     0.90 &     0.88 &     0.84 &     0.51 &     0.66 &     \textbf{0.36} &     \textbf{0.34} &     \textbf{0.46} &     0.81 &     0.84 &     0.96 &    \textbf{-0.09} \\ 
South &     0.74 &     0.87 &     0.76 &     0.76 &     0.66 &     \textbf{0.32} &     \textbf{0.41} &     \textbf{0.34} &     0.63 &     0.65 &     0.63 &     0.87 &     \textbf{0.02} \\ 
Southeast &     0.90 &     0.91 &     0.91 &     0.82 &     0.87 &     0.85 &     0.76 &     \textbf{0.45} &     0.74 &     0.88 &     0.95 &     0.95 &     0.67 \\ 
Northeast &     0.67 &     0.94 &     0.91 &     0.88 &     0.72 &     0.83 &     0.65 &     0.52 &     0.62 &     0.75 &     0.87 &     0.91 &     \textbf{0.16} \\ 
Central &     \textbf{0.48} &     \textbf{0.42} &     0.91 &     \textbf{0.20} &     0.83 &     0.85 &     0.88 &     0.56 &     0.76 &     0.93 &     0.80 &     \textbf{0.38} &     0.62 \\ 
Upper Midwest &     0.50 &     0.71 &     0.95 &     0.79 &     0.63 &     0.77 &     0.75 &     0.72 &     0.65 &     0.92 &     0.74 &     0.61 &     0.56 \\ 
Plains &     \textbf{0.11} &     0.88 &     0.94 &     0.91 &     0.63 &     0.61 &     \textbf{0.43} &     0.74 &     0.85 &     0.59 &     0.60 &     \textbf{0.06} &     \textbf{0.11} \\ 
Southwest &     0.65 &     0.75 &     0.90 &     0.90 &     0.90 &     0.83 &     0.60 &     0.60 &     0.59 &     0.93 &     0.79 &     \textbf{0.35} &     0.74 \\ 
Northwest &     \textbf{0.40} &     \textbf{0.32} &     \textbf{0.29} &     0.61 &     0.72 &     0.90 &     0.70 &     0.94 &     0.97 &     \textbf{0.43} &     \textbf{0.36} &     0.71 &     0.60 \\ 
West &     0.84 &     0.78 &     \textbf{0.27} &     0.61 &     0.80 &     0.81 &     0.75 &     0.92 &     0.85 &     0.86 &     \textbf{0.27} &     \textbf{0.43} &     0.64 \\ 
\end{tabular}
\label{tab:neg.corr}
\end{center}
\end{table}

\begin{table}
\begin{center}
  \caption{Correlation of monthly averages of positive NLDN flash counts and
    CP. Low-skill values (less than 0.5) are indicated in bold.}
\begin{tabular}{lrrrrrrrrrrrrr}
Region & J & F & M & A & M & J & J & A & S & O & N & D & annual\\ 
\hline\hline
CONUS &     0.86 &     0.96 &     0.86 &     0.85 &     0.66 &     0.64 &     0.72 &     \textbf{0.48} &     0.80 &     \textbf{0.41} &     0.87 &     0.97 &     0.80 \\ 
South &     0.59 &     0.90 &     0.77 &     0.75 &     0.83 &     \textbf{0.49} &     \textbf{0.41} &     \textbf{0.22} &     0.70 &     \textbf{0.41} &     0.63 &     0.96 &     \textbf{0.43} \\ 
Southeast &     0.93 &     0.92 &     0.84 &     0.82 &     0.83 &     0.50 &     0.56 &     \textbf{0.41} &     0.59 &     0.56 &     0.90 &     0.93 &     0.67 \\ 
Northeast &     0.80 &     0.96 &     0.92 &     0.96 &     0.64 &     0.93 &     0.80 &     0.66 &     0.82 &     0.71 &     0.97 &     0.98 &     0.87 \\ 
Central &     0.61 &     \textbf{0.31} &     0.88 &     \textbf{0.25} &     0.89 &     0.92 &     0.88 &     0.81 &     0.98 &     0.88 &     0.77 &     \textbf{0.40} &     0.89 \\ 
Upper Midwest &     \textbf{0.32} &     0.79 &     0.82 &     0.95 &     0.91 &     0.92 &     0.81 &     0.72 &     0.90 &     0.90 &     0.74 &     \textbf{0.49} &     0.81 \\ 
Plains &     \textbf{0.25} &     0.95 &     0.92 &     0.96 &     0.84 &     0.79 &     0.88 &     0.80 &     0.90 &     \textbf{0.40} &     0.67 &     \textbf{0.17} &     0.80 \\ 
Southwest &     0.76 &     0.79 &     0.96 &     0.92 &     0.84 &     0.64 &     0.50 &     0.82 &     0.50 &     0.91 &     0.79 &     \textbf{0.15} &     0.59 \\ 
Northwest &     0.88 &     0.57 &     0.71 &     0.60 &     0.59 &     0.90 &     0.83 &     0.93 &     0.97 &     0.73 &     0.65 &     0.88 &     0.67 \\ 
West &     0.86 &     0.81 &     \textbf{0.47} &     0.62 &     0.66 &     0.94 &     0.88 &     0.86 &     0.84 &     0.89 &     0.58 &     0.51 &     0.64 \\ 
\end{tabular}
\label{tab:pos.corr}
\end{center}
\end{table}

%

\end{document}